\documentclass[conference]{IEEEtran}
\IEEEoverridecommandlockouts
\usepackage{amsmath,amssymb,amsfonts}
\usepackage{algorithmic}
\usepackage{graphicx}
\usepackage{textcomp}
\usepackage{xcolor}
\usepackage{multirow}
\usepackage{pifont}
\usepackage{tabularx}
\usepackage{multirow}
\usepackage[numbers,sort&compress]{natbib}
\begin{document}


\title{TextDiffSeg: Text-guided Latent Diffusion Model for 3d Medical Images Segmentation}

\author{{Kangbo Ma}\\{mkbbupt@bupt.edu.cn}}

\maketitle

\begin{abstract}
Diffusion Probabilistic Models (DPMs) have demonstrated significant potential in 3D medical image segmentation tasks. However, their high computational cost and inability to fully capture global 3D contextual information limit their practical applications. To address these challenges, we propose a novel text-guided diffusion model framework, TextDiffSeg. This method leverages a conditional diffusion framework that integrates 3D volumetric data with natural language descriptions, enabling cross-modal embedding and establishing a shared semantic space between visual and textual modalities. By enhancing the model's ability to recognize complex anatomical structures, TextDiffSeg incorporates innovative label embedding techniques and cross-modal attention mechanisms, effectively reducing computational complexity while preserving global 3D contextual integrity. Experimental results demonstrate that TextDiffSeg consistently outperforms existing methods in segmentation tasks involving kidney and pancreas tumors, as well as multi-organ segmentation scenarios. Ablation studies further validate the effectiveness of key components, highlighting the synergistic interaction between text fusion, image feature extractor, and label encoder. TextDiffSeg provides an efficient and accurate solution for 3D medical image segmentation, showcasing its broad applicability in clinical diagnosis and treatment planning.

\end{abstract}


\section{Introduction} 
Volumetric medical image segmentation, aimed at extracting 3D regions of interest such as organs, lesions, and tissues, is a cornerstone in medical image analysis. By leveraging volumetric data from imaging modalities like CT and MRI, this task enables precise modeling of the 3D structural information of the human body, which is critical for clinical diagnosis, treatment planning, and disease monitoring. Compared to 2D medical image segmentation \cite{DBLP:conf/miccai/ZhouSTL18,Siddique_2021,shaker2024unetrdelvingefficientaccurate,cao2021swinunetunetlikepuretransformer}, volumetric segmentation presents unique challenges. The annotation process for 3D data is highly labor-intensive, requiring significant domain expertise, while the computational demands for processing volumetric data are substantial. 

Traditional approaches to 3D medical segmentation predominantly rely on encoder-decoder architectures, exemplified by U-Net and its numerous variants\cite{cicek20163dunetlearningdense,hatamizadeh2022swinunetrswintransformers,lee20233duxnetlargekernel,tang2022selfsupervisedpretrainingswintransformers}. These architectures utilize skip connections to integrate multi-scale features and have demonstrated promising results. Nevertheless, convolutional neural network (CNN)-based architectures\cite{9578194} are inherently constrained by their limited receptive fields, which restrict their ability to capture global contextual information—a critical factor for accurately segmenting complex anatomical structures.

In recent years, diffusion models\cite{ho2020denoisingdiffusionprobabilisticmodels} have emerged as a transformative approach in computer vision, excelling in tasks such as image generation\cite{zhang2023surveydiffusionbasedimage,rombach2022highresolutionimagesynthesislatent} and restoration\cite{li2023diffusionmodelsimagerestoration}. Denoising Diffusion Probabilistic Models (DDPMs), as a representative example, have been adapted for 3D medical image segmentation, offering a probabilistic framework that iteratively refines noisy data to produce high-quality outputs\cite{10.1007/978-981-96-2644-1_1,Ding_2024,tursynbek2023unsuperviseddiscovery3dhierarchical}. Diffusion-based methods have proven effective for segmenting various organs in CT and MRI scans, such as the liver\cite{christ2017automaticlivertumorsegmentation} and abdomen\cite{BENHAMIDA2021104730}. Their ability to handle complex shapes and small regions ensures high segmentation accuracy in many scenarios. However, the high dimensionality of 3D data necessitates extensive network architectures to capture global contextual information, resulting in substantial computational overhead. Latent Diffusion Models\cite{rombach2022highresolutionimagesynthesislatent} addressed this challenge by introducing VAEs to efficiently reduce data dimensions while preserving essential features. Taking inspiration from this dimensional reduction approach, studies \cite{zaman2025latentdiffusionmedicalimage} and \cite{10651343} have adapted similar latent space techniques for medical image segmentation tasks.

In addition, to mitigate computational complexity, existing approaches frequently employ 2D slices or sliding local 3D patches as inputs\cite{bieder2024memoryefficient3ddenoisingdiffusion,choo2024sliceconsistent3dvolumetricbrain}. While these strategies reduce computational demands, they inevitably compromise the structural integrity of volumetric data, leading to diminished segmentation performance. Concurrently, some researches\cite{gao2024refsam3dadaptingsamcrossmodal,feng2024enhancinglabelefficientmedicalimage,9710099} intergrate medical textual information into diffusion model frameworks provides supplementary semantic context, thereby reducing dependence on extensive pixel-level annotations. 

To address these challenges, we propose TextDiffSeg. Motivated by the limitations of existing methods in capturing global 3D contextual information and their reliance on purely visual features, TextDiffSeg introduces a cross-modal approach that integrates 3D volumetric data with natural language descriptions. Specifically, TextDiffSeg leverages a conditional diffusion process to iteratively refine segmentation results by incorporating both visual and textual information. To efficiently handle the computational demands of volumetric data, we introduce a 3D latent representation within the diffusion framework, which significantly reduces the processing cost while preserving global 3D contextual information. Additionally, we design a cross-modal attention mechanism that aligns textual descriptions with visual features, enabling the model to establish a shared semantic space between the two modalities. This design allows the model to effectively utilize complementary information from textual inputs, improving its ability to segment complex anatomical structures. By combining these innovations, TextDiffSeg achieves superior performance in challenging tasks such as multi-organ and tumor segmentation, while also demonstrating strong generalization across diverse datasets. The key contributions of our method include: 

\begin{itemize}
    \item We enhance segmentation accuracy by incorporating textual guidance, which provides complementary semantic information to facilitate better recognition of complex anatomical structures.
    \item We introduce a 3D latent representation within the diffusion framework for the first time, effectively reducing computational complexity while preserving global 3D contextual information, enabling efficient processing of large-scale volumetric data.
    \item TextDiffSeg improves the generalization ability of segmentation models across diverse tasks, including challenging cases such as multi-organ and tumor segmentation, by leveraging a shared semantic space between textual and visual modalities.
\end{itemize}

\section{Method}
\begin{figure*}
    \centering
    \includegraphics[width=0.9\textwidth]{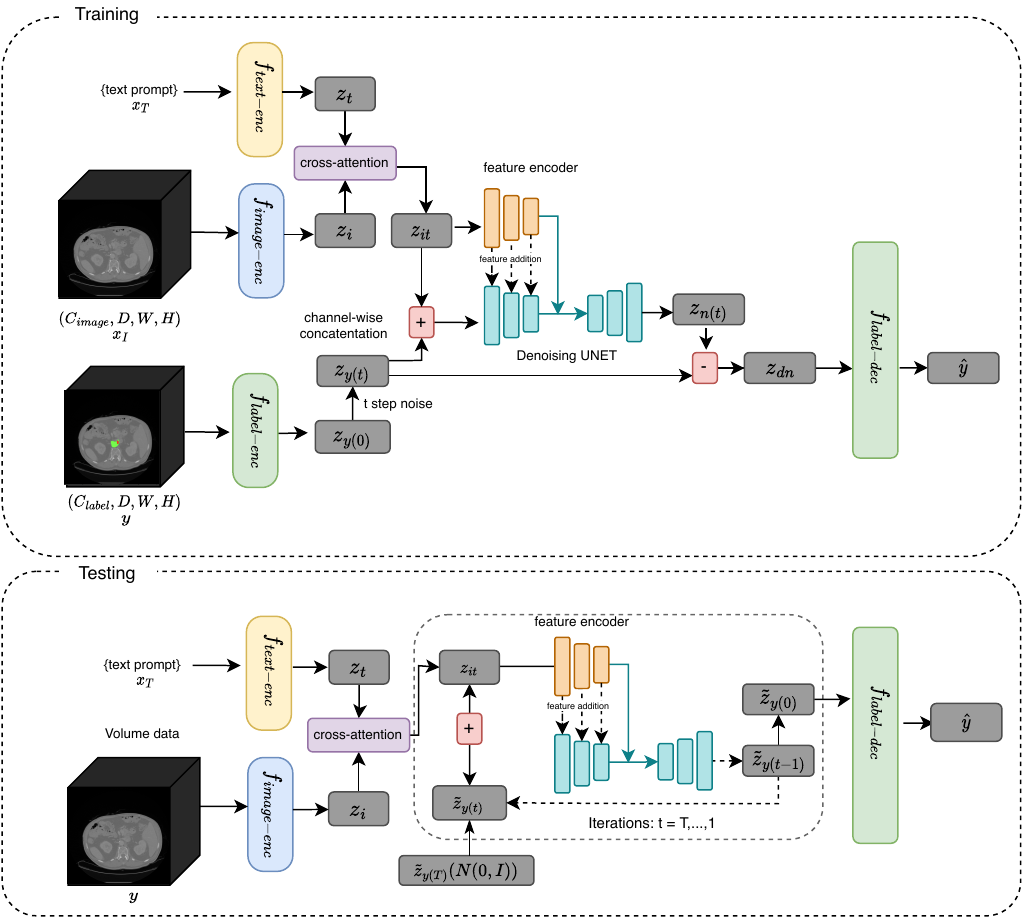}     
    \caption{The overview of TextDiffSeg consisting training phase and testing phase.}
    \label{fig:prompt-}
\end{figure*}

\subsection{Overview of TextDiffSeg}
The proposed method employs a conditional diffusion framework for 3D medical image segmentation, where the model progressively refines segmentation labels through iterative denoising.  Fig \ref{fig:prompt-} shows the training and inference process.

\subsection{Cross-modal Embedding}
Cross-modal embedding serves as a pivotal component in integrating multi-modal information for medical image segmentation tasks. By leveraging both 3D volumetric image data and natural language descriptions, this embedding establishes a shared semantic space that enables effective interaction between visual and textual modalities. The framework comprises three key elements: a 3D image encoder, a text encoder, and a cross-modal attention mechanism. The 3D image encoder extracts compact volumetric representations that capture anatomical structures and contextual information from high-dimensional medical volumes, while the text encoder generates semantic embeddings from natural language descriptions of anatomical and pathological features. These embeddings are subsequently fused through a cross-attention mechanism, which selectively aligns relevant visual features with textual context, enhancing the model's ability to focus on subtle anatomical details guided by textual cues. This unified embedding not only facilitates multi-modal understanding but also significantly improves the segmentation performance by incorporating complementary information from both modalities.

\subsubsection{Image Encoder}
The 3D image encoder, denoted by $f_{\text{3D-image-enc}}$, learns the low-dimensional volumetric embedding $z_i$ from the source 3D medical volume $x \in \mathbb{R}^{C \times D \times H \times W}$. This encoder transforms the high-dimensional input volume into a compact representation $z_i \in \mathbb{R}^{c \times d \times h \times w}$, where $c$ represents the feature channel dimension, and $d, h, w$ are the downsampled spatial dimensions ($d \ll D$, $h \ll H$, $w \ll W$). The resulting embedding captures essential anatomical structures and contextual information across the entire volume while significantly reducing the computational and memory requirements for the subsequent diffusion process.

\subsubsection{Text Encoder}

The text encoder, denoted by $f_{\text{text-enc}}$, learns the semantic embedding $z_t$ from the natural language description $T$ of anatomical structures and pathological features. Specifically, given a text annotation $t$, the text embedding process can be formulated as:
\begin{equation}
    z_t = f_{\text{text-enc}}(t) = \mathcal{E}_{\text{te}}(t)
\end{equation}

where $\mathcal{E}_{\text{te}}$ represents the BioBERT backbone pre-trained on MIMIC III dataset for obtaining clinical-aware text embeddings, and $z_t$ is the resulting text feature embedding that will be used in the subsequent cross-modal attention mechanism.

\subsubsection{3D Cross-modal Attention}
We employ a cross-attention mechanism to fuse 3D image features $z_i \in \mathbb{R}^{c \times d \times h \times w}$ with text embeddings $z_t \in \mathbb{R}^{d_t}$. First, we reshape the voxel features into sequence form $z_i' \in \mathbb{R}^{(d \times h \times w) \times c}$ as queries, while the text embeddings are linearly projected to generate key-value pairs. The cross-attention is computed as:

\begin{equation}
    z_{\text{fused}} = z_i + \text{reshape}\left(\text{softmax}\left(\frac{z_i' W_q (z_t W_k)^T}{\sqrt{d_k}}\right) z_t W_v\right)
\end{equation}

where $W_q$, $W_k$, $W_v$ are learnable parameter matrices, and $\sqrt{d_k}$ is a scaling factor. This mechanism enables the model to selectively focus on relevant anatomical structures based on textual descriptions, enhancing the segmentation model's ability to recognize subtle anatomical features.

\subsection{Label Embedding}
We note that segmentation labels in 3D medical images are discrete, and hence corrupting them by Gaussian noise is unnatural, as the volumetric label/mask has only a few modes (i.e., the number of object classes). This problem is even more pronounced in 3D, where the high dimensionality of volumetric data further complicates the application of diffusion models. We propose to mitigate this inherent problem by learning a low-dimensional standardized representation of the 3D label volumes.

Specifically, we design a 3D shape-aware label encoder $f_{\text{3D-label-enc}}(\cdot)$ that projects the input 3D labels into a continuous latent space. This encoder employs a lightweight 3D convolutional network to learn compressed shape manifolds $z \in \mathbb{R}^{k \times d \times h \times w}$, where $k \ll N$ represents the channel dimension much smaller than the original number of classes, and $d, h, w$ represent the downsampled spatial dimensions.

After obtaining the initial label embedding $z_l(0)$ from the label encoder, we apply a forward diffusion process to gradually add noise. The noisy label embedding at timestep $t$ is defined as:
\begin{equation}
    z_l(t) = \sqrt{\bar{\alpha}_t} z_l(0) + \sqrt{1 - \bar{\alpha}_t} \, \epsilon, \quad \epsilon \sim \mathcal{N}(0, I)
\end{equation}

where $\bar{\alpha}_t = \prod_{i=1}^t \alpha_i$ represents the cumulative product of noise scheduling coefficients, and $\epsilon$ is standard Gaussian noise.

During training, we sample random timesteps $t$ and optimize the denoiser network to predict the original noise $\epsilon$ added to $z_l(0)$. 


\subsection{Conditional Denoising Module}
The standard denoising mechanism in Denoising Probabilistic Models (DPMs) is designed to take two inputs: a noisy version of the input image and the corresponding timestep. However, for segmentation tasks, additional conditioning information is required to guide the denoising process. In this study, we introduce cross-modal embeddings as the conditioning input for the denoiser, ensuring that the embedding size is consistent with that of the label embeddings. Specifically, the cross-modal embedding is concatenated with the noisy representation of the label embedding to form a dual-channel input, while the timestep information is provided as a separate input.

The denoiser, denoted as $f_{\text{denoiser}}(\cdot)$, is trained to capture the transitional noise distribution of the label embedding, conditioned on the cross-modal embedding, and to predict the noise corresponding to a given timestep. To translate the denoised latent representation back into the semantic segmentation space of the original image domain, we further employ a label decoder, denoted as $f_{\text{label-dec}}$, which is trained to map the latent representation to the final segmentation output.

\subsection{Loss Function}
 The proposed loss function is designed to learn the conditional probability distribution $ q(y|X) = \mathbb{E}_{q_t(z_t|X)} [q_s(y|z)] $, where $ q_t(z|y,X) \sim \mathcal{N}(z_{\text{dn}}, \sigma^2I) $. It consists of two components: the segmentation loss $ L_1 $ and the denoiser loss $ L_2 $. The segmentation loss combines the cross-entropy loss and DSC loss.
\begin{equation}
L_1 = \mathbb{E}_{X, y} \left[ L_{\text{CE}}(\hat{y}, y) + \gamma L_{\text{DSC}}(\hat{y}, y) \right],
\end{equation}
where $ \gamma $ balances the two terms. The denoiser loss regularizes the latent space by encouraging the denoising network $ f_{\text{denoiser}} $ to reconstruct added Gaussian noise $ \epsilon $ from noisy latent embeddings $ z_l(t) $ and cross-modal embeddings $ z_{it} $. 
\begin{equation}
L_2 = \mathbb{E}_{\epsilon \sim \mathcal{N}(0, I)} \left[ \| f_{\text{denoiser}}(z_l(t), z_{it}, t) - \epsilon \|^2 \right]
\end{equation}
The total loss can be expressed as:
\begin{equation}
L = L_1 + \lambda L_2,
\end{equation}
where $ \lambda $ controls the influence of $ L_2 $, enables end-to-end training. 

\section{Experiment}
\begin{figure*}
    \centering
    \includegraphics[width=0.9\linewidth]{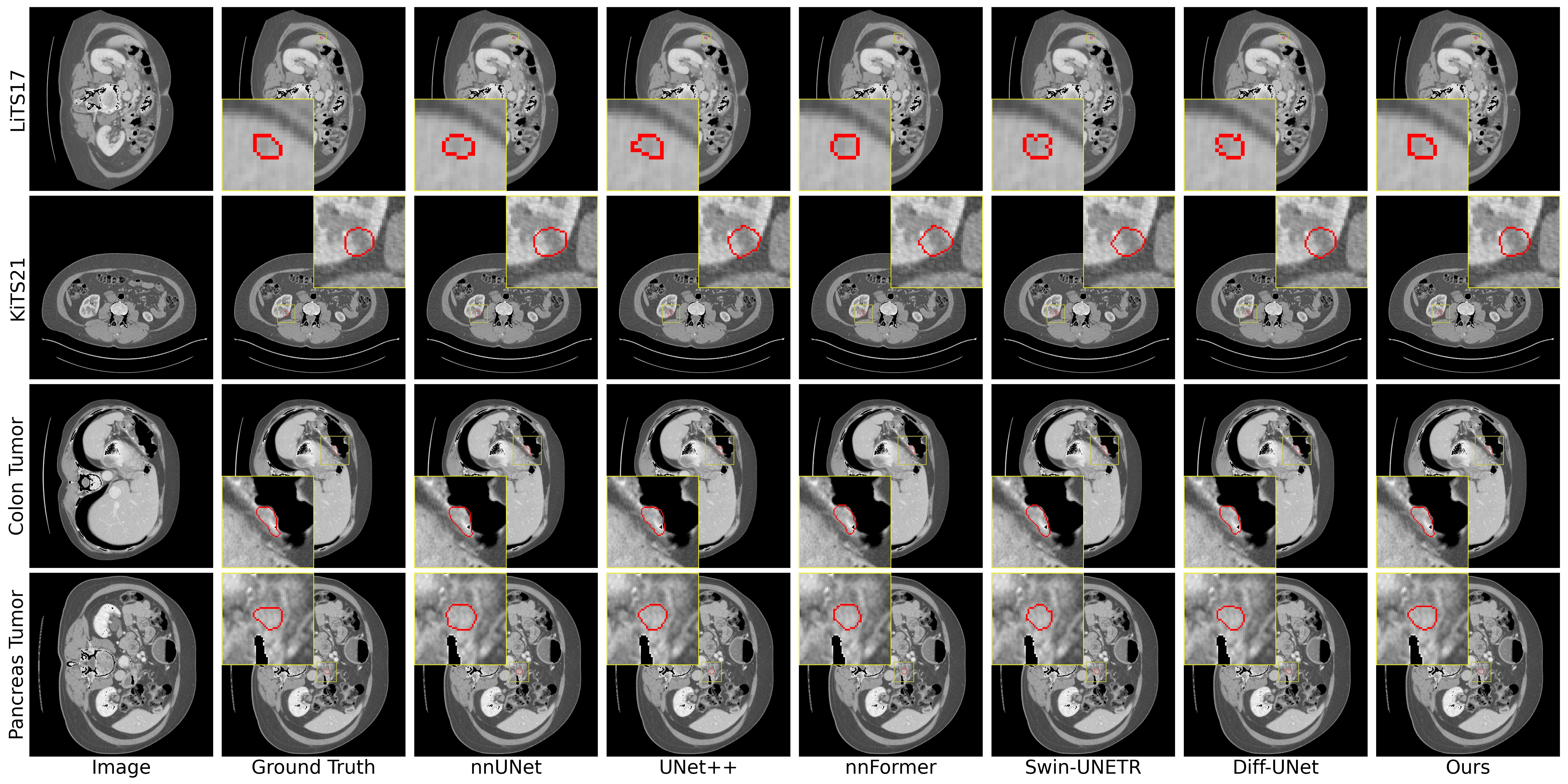}
    \caption{Qualitative visualizations of our method and baseline approaches on liver tumor, kidney tumor, pancreas tumor and colon cancer segmentation tasks.}
    \label{fig:experiment}
\end{figure*}

\subsection{Dataset}
To evaluate the volumetric segmentation performance of our method, we utilize five publicly available medical image segmentation datasets, including KiTS21\cite{heller2023kits21challengeautomaticsegmentation}, MSD Pancreas\cite{Wu2023Medical}, LiTS \cite{Bilic_2023} and MSD-Colon\cite{Antonelli2022Medical} datasets. We adopted the Dice coefficient (DICE) and Normalized Surface Distance (NSD) as evaluation metrics for quantitative comparison across all datasets.

\textbf{KiTS21 Dataset:} The KiTS21 dataset\cite{heller2023kits21challengeautomaticsegmentation}is designed for the segmentation of kidneys, tumors, and cysts in CT imaging. It comprises 300 publicly available training cases and 100 withheld testing cases. The dataset is formatted in 3D CT with files stored in .nii.gz format. The image dimensions exhibit significant variability, with spacing ranging from (0.5, 0.44, 0.44) to (5.0, 1.04, 1.04) mm and sizes ranging from (29, 512, 512) to (1059, 512, 796) voxels. All training cases contain annotations for kidneys and tumors, with cysts appearing in 49.33\% of the cases.

\textbf{MSD Pancreas Dataset:} The MSD Pancreas dataset\cite{Wu2023Medical} consists of 281 contrast-enhanced abdominal CT scans with annotations for both the pancreas and pancreatic tumors. Each CT volume has a resolution of $512\times512$ pixels, with the number of slices per scan ranging from 37 to 751. Following previous studies, we merged the pancreas and pancreatic tumor masks into a single entity for segmentation.

\textbf{LiTS Dataset} The LiTS dataset\cite{Bilic_2023} contains 201 abdominal CT scans focused on liver and liver tumor segmentation. The dataset is divided into 131 training cases and 70 testing cases. The resolution and quality of the CT images vary, with axial resolutions ranging from 0.56 mm to 1.0 mm and z-direction resolutions ranging from 0.45 mm to 6.0 mm.

\textbf{MSD-Colon Dataset} The MSD-Colon dataset\cite{Antonelli2022Medical} includes 190 abdominal CT scans, divided into 126 training cases and 64 testing cases. Each case is annotated with segmentation masks identifying the primary colon cancer regions.

\subsection{Implentation Details}

Our network is implemented in PyTorch and experiments were conducted on NVIDIA RTX A6000 GPUs.  In the training phase, each iteration involves random sampling of n patches of size $96\times 96 \times 96$, which are augmented with random flips, rotations, intensity scaling, and shifts to enhance model robustness. We employ the AdamW optimizer with a weight decay of $1e-5$, and the learning rate is initially set to $1e-4$. A linear warmup period—either set as $1/10$ of the total epochs or 30 epochs is used before applying a Cosine Annealing schedule for further adjustments.  The network architecture features a Denoising Module that comprises a Denoising UNet and a Feature Encoder, both constructed based on the framework described in [46]. The label encoder is enhanced with a final normalization layer to ensure that the label embeddings follow a standardized distribution $(\mu = 0, \sigma = 1)$. In parallel, the final two down-sampling layers of the image encoder incorporate multi-head attention layers [28] to capture more robust imaging features. The denoiser is designed with a standard ResUnet architecture, enriched by time-embedding blocks and self-attention layers, and accepts a two-channel input that fuses the cross-modal embedding with the noisy label representation, alongside its corresponding timestep. The decoder, resembling that of a typical ResUnet but without skip connections, concludes with a softmax activation layer to generate the probabilistic distribution over different object classes.  During testing, the model uses a DDIM sampling strategy with $10$ sampling steps, and each sample maintains the $96\times 96 \times 96$ dimension. A sliding window approach with an overlap rate of $0.5$ is employed to ensure the entire volume is accurately predicted.

\begin{table*}[t!]
\centering
\label{TABNLE1}
\caption{Comparison with classical medical image segmentation methods on four tumor segmentation datasets.}
\renewcommand\arraystretch{1.5}
\centering
\setlength{\tabcolsep}{4.18mm}{
\begin{tabular}{c|cc|cc|cc|cc}
\hline
\multirow{2}{*}{Methods} & \multicolumn{2}{c|}{Kidney Tumor} & \multicolumn{2}{c|}{Pancreas Tumor} & \multicolumn{2}{c|}{Liver Tumor} & \multicolumn{2}{c}{Colon Cancer} \\ \cline{2-9} 
& DICE↑    & NSD↑            & DICE↑            & NSD↑             & DICE↑           & NSD↑           & DICE↑           & NSD↑           \\ \hline
UNETR++                  & 56.49           & 60.04           & 37.25            & 53.59            & 37.13           & 51.99          & 25.36           & 30.68          \\
Swin-UNETR               & 65.54           & 72.04           & 40.57            & 60.05            & 50.26           & 64.32          & 35.21           & 42.94          \\
nnU-Net                  & 73.07           & 77.47           & 41.65            & 62.54            & 60.10           & 75.41          & 43.91           & 52.52          \\
3D U-Net                 & 78.93           & 83.13           & 55.29            & 72.80            & 63.32           & 75.41          & 50.67           & 64.71          \\
Diff-Unet                & 80.23           & 84.79           & 60.32            & 78.13            & 71.37           & 82.14          & 55.32           & 70.32          \\
TextDiffSeg  & \textbf{88.31}          & \textbf{91.45} & \textbf{71.88} & \textbf{89.91}  & \textbf{84.47}  & \textbf{93.79}  & \textbf{75.62}   & \textbf{86.16}  \\ \hline
\end{tabular}}
\end{table*}

\subsection{Comparison with SOTA Methods}

To validate the effectiveness of TextDiffSeg, we conducted experiments on four tumor segmentation datasets: kidney tumor, pancreas tumor, liver tumor, and colon cancer, comparing it with several state-of-the-art (SOTA) methods, including UNETR++\cite{shaker2024unetrdelvingefficientaccurate}, Swin-UNETR\cite{hatamizadeh2022swinunetrswintransformers}, nnU-Net\cite{isensee2018nnunetselfadaptingframeworkunetbased}, 3D U-Net\cite{cicek20163dunetlearningdense}, and Diff-Unet\cite{xing2023diffunetdiffusionembeddednetwork}. The evaluation metrics used were DICE and NSD, which assess segmentation accuracy and boundary quality, respectively. Across all datasets, TextDiffSeg consistently outperformed competing methods. Figure \ref{fig:experiment} shows qualitative visualizations of theses tasks and Table I presents the experimental results of our proposed TextDiffSeg method across a diverse set of medical image segmentation tasks.

Specifically, in kidney tumor segmentation, TextDiffSeg achieved a DICE score of 88.31\% and an NSD score of 91.45\%, significantly higher than the second-best method, Diff-Unet, which scored 80.23\% and 84.79\%. The kidney tumor dataset, characterized by relatively large and well-defined tumor structures, demonstrates how TextDiffSeg effectively integrates textual guidance and 3D latent representations to achieve precise segmentation. Similarly, for pancreas tumors, which are notably smaller and more irregular in shape, TextDiffSeg achieved 71.88\% in DICE and 89.91\% in NSD, outperforming Diff-Unet by over 10 points in DICE, highlighting its robustness in handling challenging and variable anatomical structures. For liver tumor segmentation, TextDiffSeg achieved 84.47 in DICE and 93.79\% in NSD, significantly surpassing Diff-Unet's 71.37\% and 82.14\%. Liver tumors are often small and dispersed, making them difficult to segment accurately; however, TextDiffSeg's ability to preserve global 3D contextual information through its latent representation proved critical in achieving superior performance. Finally, in colon cancer segmentation, where tumor regions are often irregular and embedded within complex surrounding structures, TextDiffSeg achieved 75.62\% in DICE and 86.16\% in NSD, again outperforming all competing methods. This demonstrates the model's ability to leverage the shared semantic space between textual and visual modalities, enabling it to adapt to diverse and complex segmentation scenarios. Overall, these results highlight the versatility and generalization ability of TextDiffSeg, establishing it as a robust solution for volumetric medical image segmentation across a wide range of tumor types and anatomical challenges.

\subsection{Ablation Study}
To evaluate the effectiveness of different components within our TextDiffSeg framework, we conducted systematic ablation experiments on three critical modules: text fusion, image feature extractor, and label encoder. We designed four variants: (1) TextDiffSeg: the complete framework with all components integrated; (2) $\zeta_1$: replacing the text fusion module with direct feature concatenation; (3) $\zeta_2$: substituting the sophisticated image feature extractor with a simple downsampling operation; and (4) $\zeta_3$: replacing the label encoder with a basic dimensionality reduction technique. As shown in Table \ref{}, removing any component results or simplifying components in performance degradation, with the complete model achieving superior performance, with the most significant performance drop observed when replacing the image feature extractor $\zeta_2$, Dice decreased by 15.15\%). In addition, the absence of text fusion $\zeta_1$ leads to an 11.93\% drop in Dice score, highlighting the importance of semantic guidance in distinguishing anatomically similar structures. Without the label encoder $\zeta_3$, performance drops by 9.02\% in Dice, demonstrating its effectiveness in transforming high-dimensional discrete labels into continuous representations that enhance diffusion stability.These findings collectively demonstrate that our three modules complement each other synergistically, with the text fusion providing semantic guidance, the image feature extractor capturing spatial context, and the label encoder facilitating effective high-dimensional label modeling—all crucial for achieving state-of-the-art medical image segmentation performance.

\begin{table}[]
\caption{ABLATION ON EACH KEY COMPONENT IN OUR METHOD.}
\begin{tabular}{l|l|ll}
\hline
Model Variant & Description                              & Dice (\%) & NSD (\%) \\ \hline
TextDiff3D    & whole component & 84.47     & 93.79    \\
$\zeta_1$            & w/o text fusion module                   & 72.54     & 86.32    \\
$\zeta_2$       & replacing image encoder              & 69.32     & 81.04    \\
$\zeta_3$       & replacing label encoder                        & 75.45     & 89.41    \\ \hline
\end{tabular}
\end{table}

\section{Conclusion}
In this work, we introduced TextDiffSeg, a text-guided diffusion model framework that integrates 3D volumetric data with natural language descriptions for medical image segmentation. By addressing the limitations of traditional DPM-based methods, such as high computational costs and inadequate contextual preservation, TextDiffSeg achieves state-of-the-art performance across various segmentation tasks, including tumor and multi-organ segmentation. The proposed framework leverages cross-modal embedding, innovative label encoding, and text fusion techniques, enabling robust segmentation of complex anatomical structures while maintaining computational efficiency. Experimental results and ablation studies validate its effectiveness and highlight its potential for clinical applications, such as diagnosis, treatment planning, and personalized healthcare. TextDiffSeg paves the way for more advanced human-interactive medical imaging systems and provides a scalable solution for real-world deployment in diverse medical scenarios.


\footnotesize
\bibliographystyle{IEEEbib}
\bibliography{refs}

\vspace{12pt}
\end{document}